\documentclass[twocolumn,showpacs,preprintnumbers,amsmath,amssymb]{revtex4-1}

\usepackage{epsfig}
\usepackage{dcolumn}
\usepackage{bm}
\usepackage{xcolor}
\usepackage{amsmath}
\usepackage{float}
\usepackage{siunitx}
\usepackage{tikz}

\usepackage[normalem]{ulem}

\newcommand{\JQ}[1]{{\color{brown}{#1}}}

\begin{document}

\preprint{\today} 

\title{Numerical Studies of Ground State Atomic Parity Violation Measurements in Cesium}

\author{Chunya Wang$^{1}$, Matthew Summers$^{1}$, Jonah A. Quirk$^{1}$, Diego Garcia$^{1}$, Carol E. Tanner$^2$ and D. S. Elliott$^{1,3,4}$}

\affiliation{%
   $^1$Department of Physics and Astronomy, Purdue University, West Lafayette, Indiana 47907, USA\\
    $^2$Department of Physics and Astronomy, University of Notre Dame, Notre Dame, Indiana 46556, USA\\
      $^3$The Elmore Family School of Electrical and Computer Engineering, Purdue University, West Lafayette, Indiana 47907, USA \\
         $^4$Purdue Quantum Science and Engineering Institute, Purdue University, West Lafayette, Indiana 47907, USA\\
 }

\date{\today}

\begin{abstract}
We report results of numerical simulations of an ongoing measurement designed to measure the electric dipole (E1) transition moment between hyperfine components of the ground state of atomic cesium.  This E1 moment is non-zero solely due to nuclear weak interactions between the nucleons, leading to a nuclear anapole moment of the cesium nucleus.  Through these simulations, we explore the effect on the microwave spectrum of a static transverse magnetic field, a spatial offset of the atomic beam from the axis of the radio-frequency (rf) cavity, the angular tilt of the rf cavity, and the ac Stark shift by the Raman-coupling beams used to generate the initial superposition state between ground state components. The results of these simulations help place limits on the experimental parameters, and help guide the future experimental procedures.
\end{abstract}

\maketitle 

\section{Introduction}\label{sec:introduction}

The 1997 report by Wood {\it et al.}~\cite{WoodBCMRTW97} of measurements of atomic parity violation (APV) in cesium generated a determination of the nuclear anapole moment of the cesium nucleus.  The term `anapole moment' gained traction in 1958 with the work of Zeldovich~\cite{Zeldovich1958} who described the field distribution inside a toroidal-like current distribution within the nucleus.  This current distribution is the result of the parity-odd, time-reversal-even weak force interaction between the nucleons within the nucleus~\cite{SafronovaBDKDC2018}.  The interaction between the electron when it overlaps the nucleus and the magnetic field inside this `toroid' depends on the electronic spin, leading to a nuclear-spin dependent (NSD) interaction.  

The anapole moment value reported in the Boulder report~\cite{WoodBCMRTW97} is the only statistically significant measurement of an anapole moment in any atomic system.  This value of the anapole moment, however, is in tension with the constraints on the weak meson-nucleon coupling constants in a model introduced by Desplanques, Donoghue, and Holstein (DDH)~\cite{desplanques1980} and determined through other nuclear scattering measurements~\cite{LangMMNRSSH1985,BarrnesLDMMCAS1978,BizzetiFMPPST1980,AhrensHKMSSWBG1982,PageEEKLMMSWMMBAC1987,EvansEEKLMMMPSWMBAC1985,BiniFPT1985,AdelbergerHHWVH1983,ElsenerGKSUSFZV1984,ElsenerGKSUVSFZ1987}.  It is of great interest therefore to revisit this anapole moment measurement in order to either confirm the value reported by the Boulder group or to provide a new determination.  For this reason, we have been pursuing a new measurement of the nuclear anapole moment of cesium using a transition between hyperfine components of the ground state at 9.19 GHz~\cite{damitz2024technique}. 

The goal of the program is to measure the very weak electric dipole (E1) transition moment between the hyperfine components of the ground state of atomic cesium. This transition is very active as a magnetic dipole (M1) interaction, of course. 
The challenge is to measure this extremely weak E1 moment in the presence of the much stronger M1 moment.

In the present work, we describe a series of numerical studies that we have carried out of the excitation spectrum between the ground hyperfine levels of cesium. To set the stage for the description of the calculations, we first review, in Section \ref{sec:experiment}, the experimental procedure for the measurements.  We then discuss the numerical technique for the computations (Section \ref{sec:EoM}), followed by a discussion of the results in Section \ref{sec:results}.   The goal of these studies is to help us interpret the spectra that we have observed in previous laboratory measurements, and to guide future measurements toward meaningful results. 

\section{Description of the Measurement}\label{sec:experiment}
\begin{figure}[t!]
    \centering
    \includegraphics[width=0.9\linewidth]{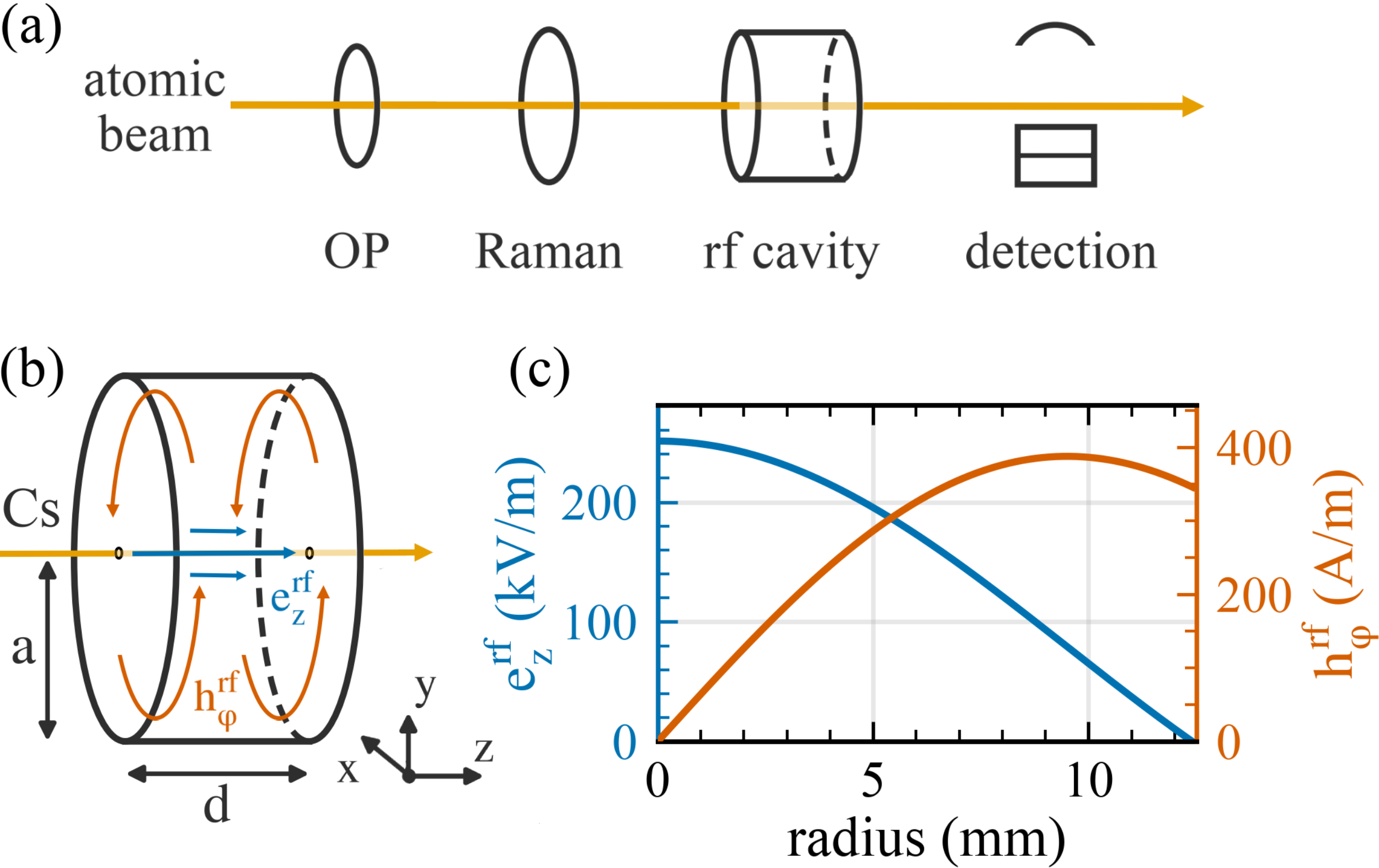}
    \caption{(a) The experimental configuration of Ref.~\cite{damitz2024technique}.  The cesium atoms in an atomic beam are first prepared by optical pumping (OP) into a single hyperfine component of the ground state by a pair of lasers tuned to hyperfine lines of the $6s \rightarrow 6p_{3/2}$ (D$_2$) line.  Using two fields tuned to the $F=3 \rightarrow F=4$ ground state Raman transition, we create a coherent superposition between the two hyperfine states. The atoms subsequently enter a cylindrical rf cavity excited on the TM$_{010}$ mode by a coherent field at 9.19 GHz.  In the detection region, we detect the incremental phase-coherent excitation of the F=4 ground state.  (b) The cylindrical cavity, showing the primary fields of the TM$_{010}$ mode. (c) Plots of the amplitudes $e^{\mathrm{rf}}_{\mathrm{z}}(\rho)$ and $h_{\phi}^{\mathrm{rf}}(\rho)$ inside the cavity vs.~the radial distance $\rho$.}
    \label{fig:exp_fig}
\end{figure}

In this section, we describe the experimental set-up used in our prior work~\cite{damitz2024technique}, setting the stage for the discussions of the simulations that follow. As shown schematically in Fig.~\ref{fig:exp_fig}(a), we generate an effusive beam of cesium atoms with a heated oven and aperture.  
As the atoms exit the oven, their population is evenly distributed among the 16 hyperfine components of the ground $6s \ ^2S_{1/2}$ state with $F=3$ and $F=4$. 
(We will use the abbreviated notation $6s$ for this state. $F$ is the total angular momentum of the atom, which includes the electronic angular momentum $J=1/2$ and nuclear spin $I=7/2$.) 
We optically pump the atoms into a single hyperfine component of the ground state ($F=3$, $m=3$, where $m$ is the quantum number for $F_z$.), prepare the ground state atoms in a coherent superposition of hyperfine components using a Raman interaction, and further drive the hyperfine transition using the fields inside a radio-frequency (rf) cavity.  We were able to demonstrate the coherence between the Raman and the subsequent rf interaction, but have not yet achieved the sensitivity required for a direct measurement of the parity-nonconserving (PNC) signal. 
The main experimental challenge was the background signal arising from magnetic dipole (M1) transitions, which were much larger than the expected PNC signal. By investigating the effects of transverse static magnetic fields, rf magnetic fields, and atomic beam misalignment, we identified a longitudinal rf magnetic field component $h_z^{\mathrm{rf}}$ of the original rf cavity as the dominant source of background. The cavity geometry and the primary field components, $e_z^{\mathrm{rf}}$ and $h_{\phi}^{\mathrm{rf}}$, of the TM$_{010}$ mode are shown in Fig.~\ref{fig:exp_fig}(b).  The field component $h_z^{\mathrm{rf}}$ is present within the cavity only because of the channels through which rf power is coupled into the cavity and holes for the atoms to enter and exit the cavity. The resulting M1 signal was approximately three to four orders of magnitude larger than the expected PNC signal. To suppress this background, we improved the rf cavity design by replacing the original two-input configuration with a four-input configuration, increasing the symmetry of the cavity. This modification significantly suppressed the unwanted longitudinal rf magnetic field component $h_z^{\mathrm{rf}}$ while enhancing the longitudinal rf electric field component $e_z^{\mathrm{rf}}$. Yet to be achieved is to precisely align the atomic beam with the axis of the rf cavity, and to reduce the transverse static magnetic fields in the rf cavity.

%

The fields driving the $F=3$ and $F=4$ ground state levels through the Raman transition are produced by narrow-band, linearly-polarized cw phase-locked optical lasers at a wavelength of $\lambda = 852$ nm, tuned close to but off-resonant from the $6s \rightarrow 6p_{3/2}$ transition (i.e.~the D$_2$ line).  (We use the abbreviated notation $6p_{3/2}$ in place of the more complete notation $6p \ ^2P_{3/2}$.)  These fields drive a $\Delta m =0$ Raman transition $6s,\ (F,m) =(3,3) \rightarrow 6s, \ (4,3)$. 
We adjust the intensities of these beams and their detuning $\Delta$ from resonance with the $6p_{3/2}$ lines to produce a rotation angle on the Bloch sphere of $\Theta_R = \Omega_R \tau_R \sim \pi/2$, where $\Omega_R$ is the Rabi precession frequency in the Raman beams and $\tau_R$ is the transit time of the atoms through the Raman beams.  For a Bloch rotation angle of $\pi/2$ the atoms are promoted to a 50:50 superposition state between the $6s, F=3$ and $6s, F=4$ states.  (Because of the broad velocity distribution of the atomic beam, only one velocity group of the atoms is truly in the target state.) 

After the atoms leave the Raman beams, they enter the cylindrical rf cavity, which is excited at the resonant frequency of the TM$_{010}$ mode at 9.19\sout{2} GHz.  The atoms traverse the rf cavity along its $z$-axis, where the rf electric field is relatively uniform and at its maximum strength and the rf magnetic field passes through a null. (See Fig.~\ref{fig:exp_fig}(c) for plots of the rf electric and magnetic field amplitudes vs.~the radial distance $\rho$.) The interaction with this rf electric field drives an additional precession angle $\Theta_{E1}$ on the Bloch sphere.  The direction of this additional precession depends on the phase difference between the rf interaction inside the cavity and the Raman interaction that prepares the 50:50 superposition state.  We modulate this phase difference at a modulation frequency of $150$ Hz, and detect the modulation of the population change in the F=4 hyperfine component of the ground state. The amplitude of this population modulation is a signature of the magnitude of the E1 transition driven by the rf field.

\section{Equations of motion of state amplitudes}\label{sec:EoM}
To model the evolution of the atom as it passes through the Raman interaction region and subsequently through the rf cavity, we numerically integrate the equations of motion of the state vector, 
\begin{equation}\label{eq:superposition}
     \psi(t)=c_g(t)\psi_ge^{-i\omega_gt}+c_e(t)\psi_ee^{-i\omega_et}.
\end{equation}
where $\psi_g$ is the $6s,\ (F,m) =(3,3)$ state, and $\psi_e$ is the $6s,\ (F,m) =(4,3)$ state.  The atoms are initially in the former.  As the atoms pass first through the Raman beam in free space, followed by their interaction with the rf field in the cavity, the evolution of the system is described by the equations of motion of the amplitudes $c_g(t)$ and $c_e(t)$, as derived from the Schr\"{o}dinger Equation~\cite{meystre2007elements,foot2005atomic} 
\begin{equation}
    \frac{dc_e(t)}{dt}=c_g(t)\left(-\frac{i}{\hbar}V e^{i\Delta \omega t}\right)
    \label{eq:equationofmotionexcited}
\end{equation}
and
\begin{equation}
    \frac{dc_g(t)}{dt}=c_e(t)\left(-\frac{i}{\hbar}(V)^*e^{-i\Delta \omega t}\right),
    \label{eq:equationofmotionground}
\end{equation}
where the interaction $V$ consists of the Raman interaction $V_{R}$ when the atoms are in the Raman beams, or the sum of a magnetic dipole $V_{\rm m}$ and weak-force-induced electric dipole $V_{\rm pnc}$ inside the cavity.   
$\Delta \omega$ is the energy difference between the states $\psi_g$ and $\psi_e$.  
We previously discussed these equations of motion in detail in Ref.~\cite{damitz2024technique}.  

The E1 interaction with the rf electric field in the cavity is 
\begin{equation}\label{eq:V_PNC}
   V_{\rm pnc} = - \boldsymbol{\mathcal{E}}_{\rm pnc} \cdot \mathbf{e}^{\rm rf},
\end{equation}
where $\boldsymbol{\mathcal{E}}_{\rm pnc}$ is the transition dipole moment that we seek to measure and $\mathbf{e}^{\rm rf}$ is the amplitude of the rf electric field.  
This field amplitude is maximal along the cavity axis and decreases off the axis as $J_0(\rho/a)$, where $J_0$ is the zero-order Bessel function, $\rho$ is the radial distance from the axis, and $a = 1.241$ cm is the radius of the cylindrical cavity. The transition dipole moment for this ground state transition has been calculated in the Random Phase Approximation (RPA)~\cite{JohnsonSS03,DzubaF12} and in the RPA with additional Brueckner-type correlations included~\cite{DzubaF12}.  The result of the latter is $\mathcal{E}_{\rm pnc} = i \kappa \ 1.82 \times 10^{-11} \ e a_0$, where $\kappa=0.117$ is a parameter that characterizes the magnitude of the nuclear-spin dependent PNC effects, including contributions from the anapole moment of the nucleus, the axial-vector nucleon and vector electron currents $(A_nV_e)$, and the combined effects of the hyperfine interaction and the vector nucleon and axial-vector electron currents $(V_nA_e)$. The value of $\kappa=0.117$ was derived from the 1997 APV measurements of Wood \emph{et al.}~\cite{WoodBCMRTW97}, as reported in Refs.~\cite{JohnsonSS03} and \cite{DzubaF12}.  Using an electric field amplitude of $e_z^{\rm rf} = 251$ kV/m, as calculated from numerical simulations of our rf cavity with 1 Watt of power delivered to each of the four inputs, we calculate a Rabi frequency $\Omega_{\rm rf} = \mathcal{E}_{\rm pnc}  e_z^{\rm rf} / \hbar = 0.0429 \ \mathrm{rad/s}$.  In the 43 $\mu$s transit time required for the peak velocity (287 m/s) atoms to travel through the cavity along the $z$-axis, the Bloch vector precesses through an angle of $\Theta_{{\rm rf}} = 1.9 \ \mu$rad.  

The interaction $V_{\rm m}$ of the atoms with the rf magnetic field $\mathbf{h}^{\rm {\rm rf}}$ is the primary source of background for the measurement of $\mathcal{E}_{\rm pnc}$, which must be highly suppressed in order to observe the PNC interaction.  We showed in Ref.~\cite{damitz2024technique} that the interaction Hamiltonian describing the magnetic dipole interaction is of the form 
\begin{equation}\label{eq:V_m2}
   V_m = -\frac{\sqrt{16-m^2}}{8} g_s \mu_B \mu_0 \:  \left\{ h_z^{\rm {\rm rf}} + \left( \frac{h_x^{\rm {\rm rf}} B_x + h_y^{\rm {\rm rf}} B_y}{B_z} \right) \right\}. 
\end{equation}
Ideally, the $h_z^{\rm rf}$-component is not present in the cavity mode as it can drive $\Delta m = 0$ transitions directly. In a perfect cylindrical rf cavity excited on the TM$_{010}$ mode, of course, the only magnetic field component is the $h_{\phi}^{\rm rf}$-component, which rotates azimuthally around the axis, is tangential to the cylindrical walls at $\rho=a$, and is zero on the cavity axis. The actual cavity, however, deviates from this ideal case in that on-axis ports are needed in the two flat faces to allow atoms to enter and exit the cavity, and side-wall ports are needed for the rf power to enter the central cavity from the outer excitation chambers (which are fed by the rf sources).  Models of the cavity field amplitudes obtained using COMSOL Multiphysics\textsuperscript{\textregistered} do indicate that the $h_z^{\rm rf}$-component is very small; as small as $\sim 2 \times 10^{-4}$ A/m in our latest simulation, and decreasing significantly with each decrease in the computational grid spacing.  Unfortunately, decreasing the grid spacing requires additional computer memory which is unavailable to us, so we can only report this value as an upper bound on $h_z^{\rm rf}$. In the computed spectra shown in the following sections, we have set the field amplitude $h_z^{\rm rf}$-component equal to zero. 

While the $h_z^{\rm rf}$-component may be sufficiently reduced, we must still deal with the $\phi$ component of the rf magnetic field. When the static transverse magnetic field components $B_x$ and $B_y$ are zero, the off-axis rf magnetic field, which has a component only in the $\phi$ direction, can drive only $\Delta m = \pm 1$ transitions, and does not contribute to the background. But when $B_x$ and $B_y$ are not zero, these field components mix projection states through the Zeeman interaction, allowing $\Delta m = 0$ transitions as well. This background is suppressed partially in that the $h_{\phi}^{\rm rf}$ component is small near the cavity axis, following a $J_1(\rho/a)$ profile, where $J_1$ is the first-order Bessel function of the first kind.  $J_1(\rho/a)=0$ at $\rho = 0$, so $h_{\phi}^{\rm rf}$ is small near the cavity axis where the atoms are.  But the atom beam has a finite width, so even a small $h_{\phi}^{\rm rf}$ can be important when the static transverse magnetic field components $B_x$ and $B_y$ are not fully suppressed. Fortunately, an additional reduction is available when the atom beam is centered on the cavity axis.  Since the rf magnetic field points in the $\phi$ direction, $h_x^{\rm rf}$ and $h_y^{\rm rf}$ are odd about the $z-$axis.  Therefore, if the atom beam is perfectly centered on the cavity axis, population transferred due to the $V_m$ interaction averages to zero. Since this interaction also depends on the state mixing by the transverse static magnetic field $B_x$ and $B_y$, the key to the measurement of $V_{\rm pnc}$ is to center the rf cavity on the atom beam, and to reduce the static transverse magnetic field components sufficiently to suppress the average value of $V_m$ to a value much less than $V_{\rm pnc}$.

\begin{figure}[t!]
  \begin{center}
     \includegraphics[width=0.4\textwidth]{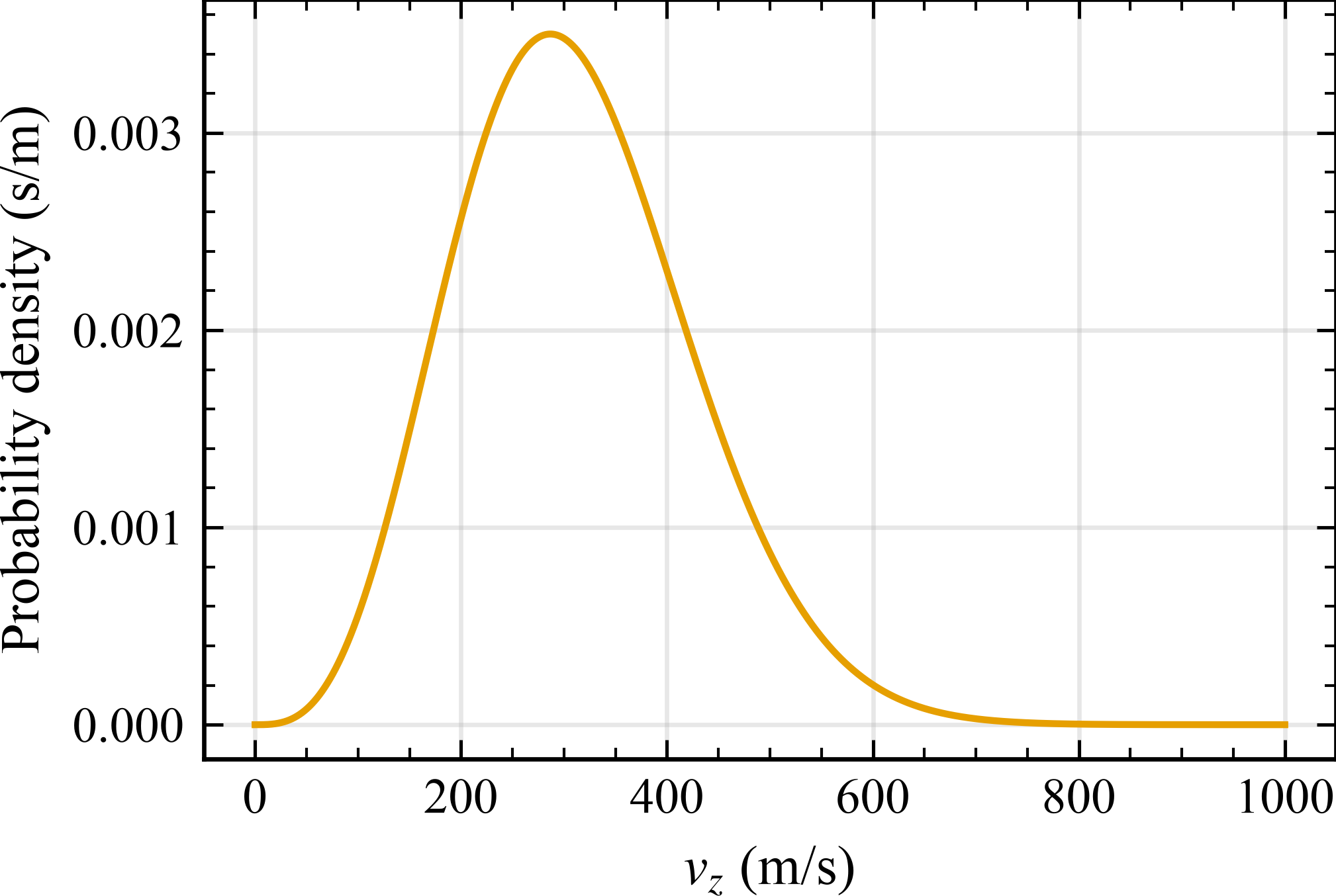}
  \caption{Velocity distribution of the cesium atoms along the propagation axis ($z$). 
  The curve shows the probability density of the effusive-beam $f(v)=2\beta^2v^3e^{-\beta v^2}$, where $\beta=m/(2k_BT)$ and T = 443 $\mathrm{K}$~\cite{ramsey1956molecular}.  The most probable and mean velocities are approximately 287 and 311 $\mathrm{m/s}$, respectively.}
  \label{fig:vz_distribution}
  \end{center}
\end{figure}

We calculate individual spectra of the interfering population modulation signal, $\Delta P$, by following the evolution of individual atoms using Eqs.~(\ref{eq:equationofmotionexcited}) and (\ref{eq:equationofmotionground}) as they travel through the Raman beams and then the interaction region.  The velocity $\mathbf{v}$ (magnitude and direction) and initial position $(x,y)$ of each atom are chosen randomly from the distribution for the apertured thermal atomic beam using a Monte Carlo technique.  
We show a calculated distribution of velocities~\cite{ramsey1956molecular} in Fig.~\ref{fig:vz_distribution}.  
The atomic velocity distribution determines the interaction time of the atoms with the Raman and rf fields and is essential for modeling the measured spectrum. This figure shows the velocity distribution before all the atoms were selected by the collimator. 

We follow the evolution of 20,000 atoms to generate each data point within a spectrum.  
Considering only a single atomic velocity and an ideal trajectory cannot fully capture the behavior of the atomic ensemble under realistic experimental conditions. Monte Carlo simulation allows us to predict the behavior of a thermal atomic beam by averaging the quantum dynamics of individual atoms. The Monte Carlo method incorporates all effects into a unified numerical model by averaging the independently calculated evolution of a large number of atoms, providing a more realistic prediction of the experimental signal. By accounting for the thermal velocity distribution, beam collimation, gravitational deflection, and spatial inhomogeneity of the rf field, this approach provides a reliable basis for evaluating systematic errors in the experiment.
We vary the phase difference $\Delta \varphi$ between the Raman and rf interactions, leading to a sinusoidal variation of the incremental population in the $F=4$ state $|c_e|^2$.  We compute the amplitude of this modulation $\Delta P$ as half the difference between the maximum and minimum values of $|c_e|^2$.

In the following section, we present a series of spectra that help place limits on the precision required to suppress the $V_m$ interaction to a value much less than $V_{\rm pnc}$.


\section{Results}\label{sec:results}
In the following, we present several series of calculated spectra of $\Delta P$ versus the detuning $\Delta$ under a variety of conditions, as well as the dependence of $\Delta P$ on different experimental parameters at zero detuning ($\Delta=0$). These calculations allow us to study the effect of an external transverse magnetic field as well as both linear and angular displacements of the rf cavity with the goal of which to determine the limits on these parameters required to sufficiently reduce systematic errors. We also explore the effect of the ac Stark shift by the Raman coupling lasers on the spectrum of $\Delta P$.

\subsection{Effect of transverse magnetic field and spatial offset of the atomic beam}
The atomic beam passes through the cylindrical rf cavity along its axis and interacts with the rf field within the cavity. The relative displacement between the cavity and the atomic beam changes the rf field experienced by the atoms (particularly the rf-magnetic field $h_{\phi}^{\rm rf}$) and thus affects the resulting spectra. Among these displacements, transverse displacement has the most noticeable effect on the interaction between the atoms and the rf field.

\begin{figure}[t!]
  \begin{center}
     \includegraphics[width=0.45\textwidth]{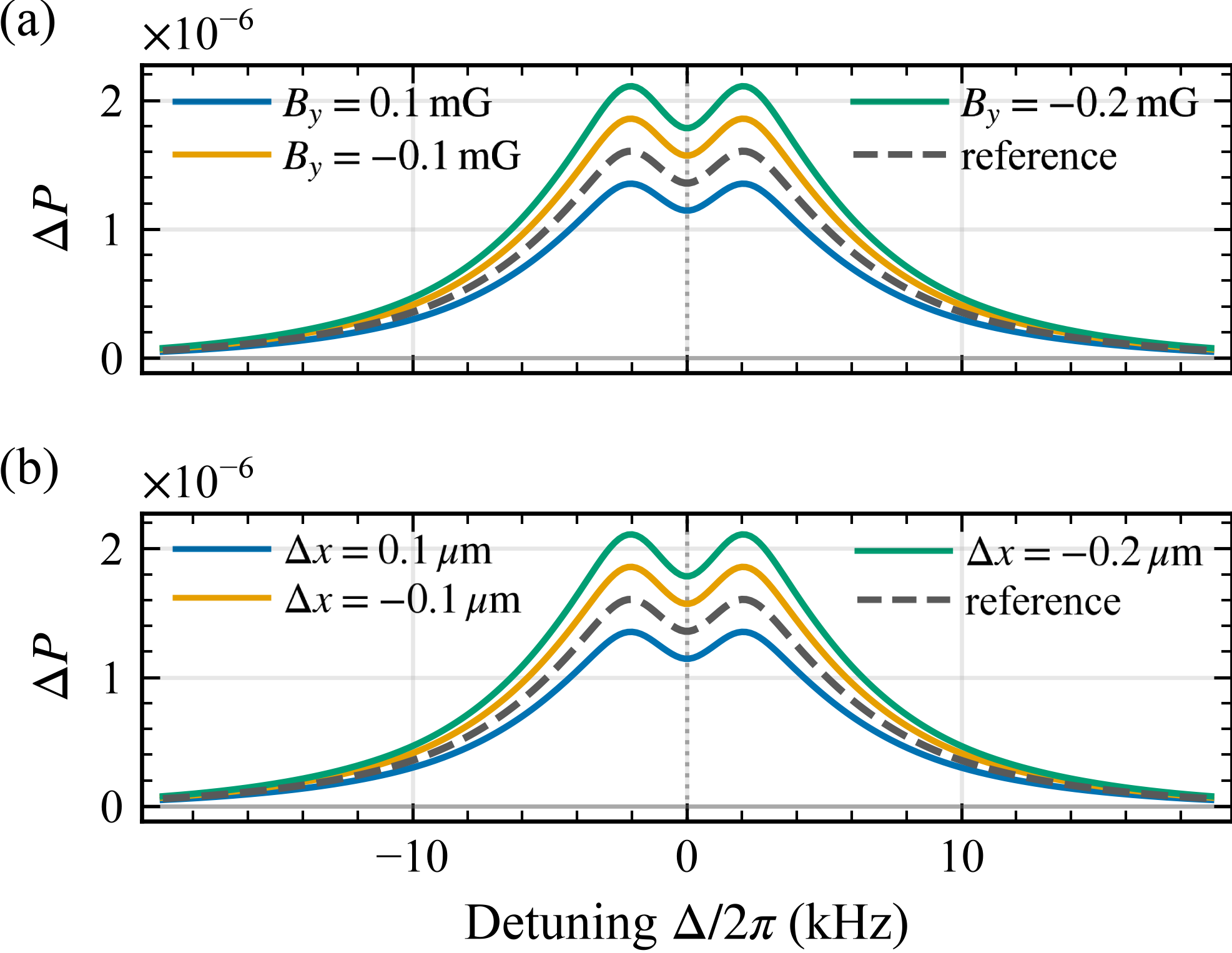}
  \caption{Spectra of the amplitude $\Delta P$ of the population modulation as a function of $\Delta/2 \pi$, the detuning of the Raman and rf field frequency from resonance. In (a), the four lines represent the spectra for varying values of the static transverse vertical magnetic field component $B_y$.  The center of the atomic beam is shifted horizontally from the cavity axis by $0.1 \, \mu$m for each of these spectra. In (b), the lines show the spectra at different transverse horizontal displacement with the magnetic field $B_y$ fixed at $0.1 \,$mG. The dashed spectrum, labeled reference, corresponds to zero transverse magnetic field and zero displacement throughout this paper.}
  \label{fig:spectra_var_B_var_x}
  \end{center}
\end{figure}

In Fig.~\ref{fig:spectra_var_B_var_x}, we show several spectra of the amplitude $\Delta P$ of the population modulation as a function of $\Delta/2\pi$, the detuning of the Raman and rf field frequency from resonance.   

For these different spectra in Fig.~\ref{fig:spectra_var_B_var_x}(a), the only difference is the value of the vertical static transverse magnetic field $B_y$ applied to the atoms. The horizontal offset $\Delta x$ for each of these spectra is fixed, equal to $0.1 \ \mu$m.   We see very little variation in the shape or width of these spectra. However, there is considerable variation in the amplitude of the peaks.  Note also that the effect of this transverse field can be to increase or decrease the amplitude of the spectrum, depending on the sign of $B_y$. Series of spectra of $\Delta P$ versus $\Delta/2\pi$ at different horizontal offsets $\Delta x$ between the atom beam and the axis of the cavity for fixed magnetic field $B_x=0.1$ mG show similar behavior, as seen in Fig.~\ref{fig:spectra_var_B_var_x}(b).

This behavior is completely consistent with the interaction term $V_m$ given in Eq.~(\ref{eq:V_m2}), in which one term of the interaction appears as the product $h_y^{\rm rf} B_y$.  Recall that $h_y^{\rm rf}$ grows with a displacement $\Delta x$ of the atomic beam in the $x$-direction. From the same equation, we can derive that the spectra show the same behavior for the other set of variables, $h_x^{\rm rf} B_x$. It is worth noting that in the vertical direction, even when the atomic beam has no offset from the cavity axis in the simulation, the resulting spectra still exhibit a small shift relative to the reference due to gravity. In the experiment, some atoms may fail to enter or exit the rf cavity due to gravity, and may therefore reduce the gravitational effect on the measured signal.

\begin{figure}[t!]
  \begin{center}
     \includegraphics[width=0.45\textwidth]{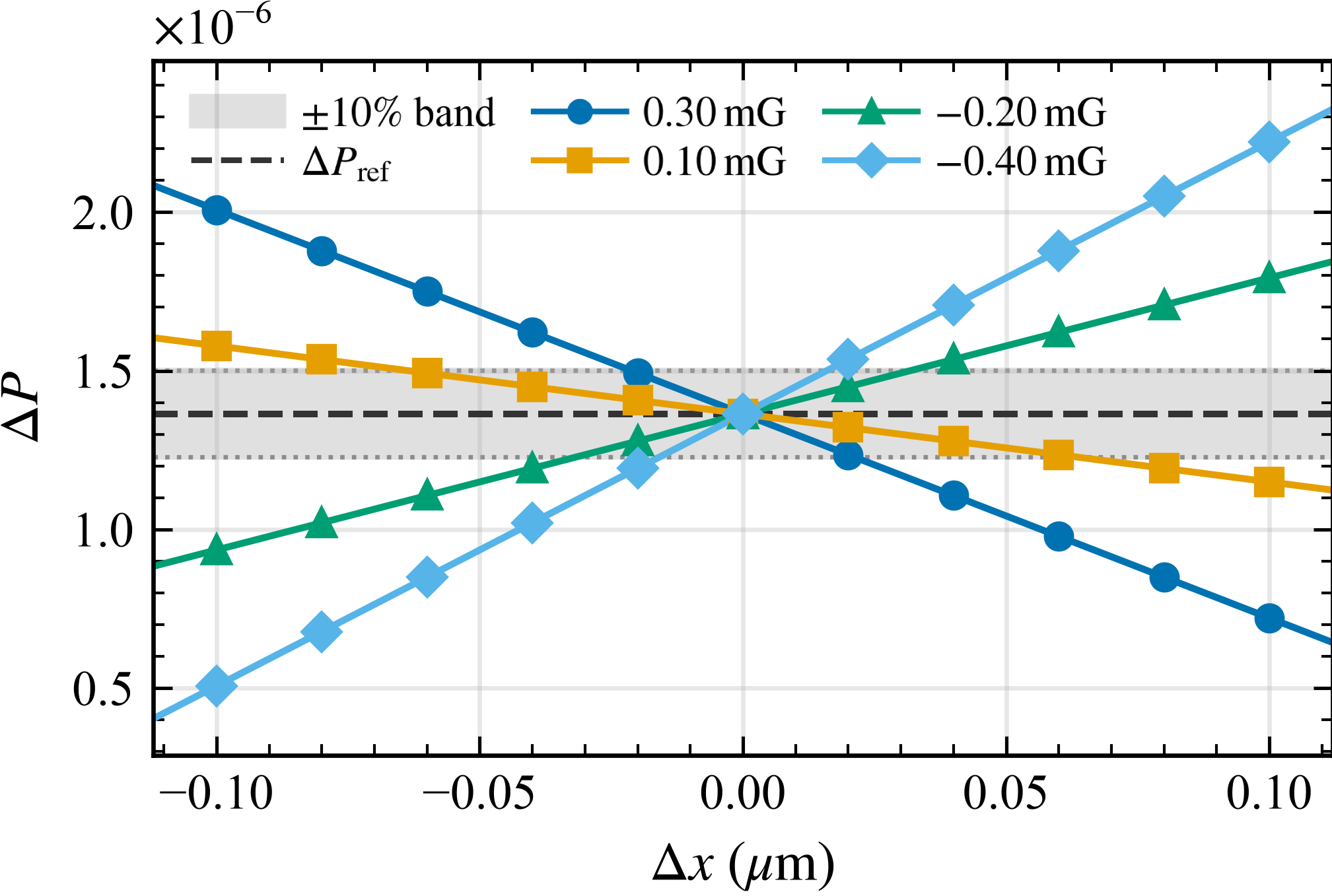}
  \caption{Plots of the amplitude $\Delta P$ of the population modulation spectrum at zero detuning $\Delta=0$ as a function of the transverse displacement. The different lines represent the amplitudes for varying values of the static transverse magnetic field component $B_y$. All the lines intersect at a common point, a feature that can be used to help align the axis of the rf cavity with the center of the atomic beam.}
  \label{fig:amp_vs_By}
  \end{center}
\end{figure}

To further illustrate the variation in amplitudes of the population modulation, we show in Fig.~\ref{fig:amp_vs_By} a series of plots of the amplitude $\Delta P$ of the population modulation at line center (i.e.\  $\Delta =0$) as a function the offset $\Delta x$ between the center of the atomic beam and the axis of the rf cavity.  Notice that the plots for different values of $B_y$ cross each other at $\Delta x =0$.  This feature will be instrumental in the measurements as we seek to align the axis of the rf cavity with the center of the atomic beam. Similarly, the lines corresponding to different $B_x$ values intersect near $\Delta y = 0$, but slightly offset (by $\sim2.3 \, \rm{nm}$) due to the effect of gravity on atom trajectories.

\begin{figure}[b!]
  \begin{center}
     \includegraphics[width=0.45\textwidth]{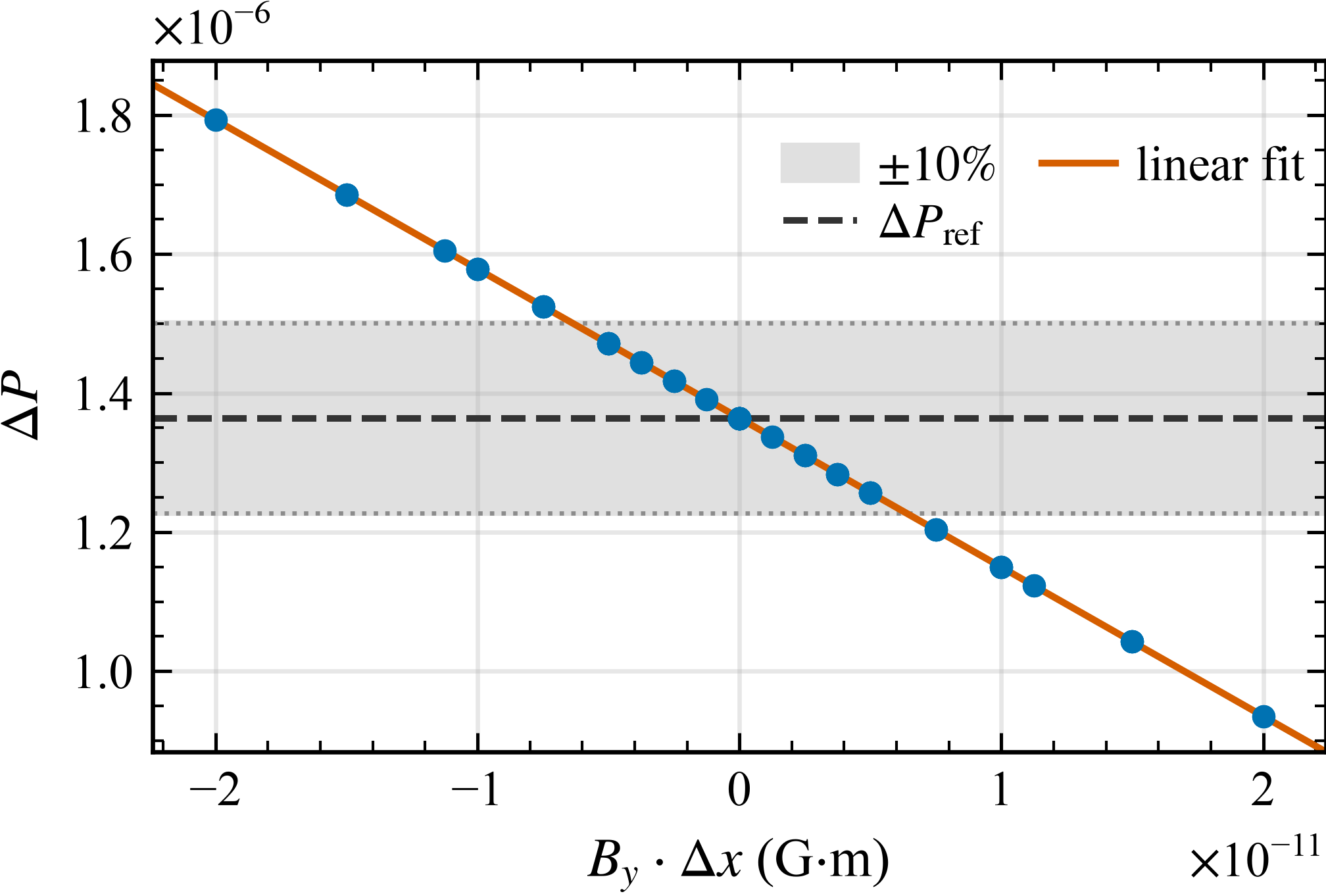}
  \caption{Population modulation amplitude $\Delta P$ at zero detuning ($\Delta=0$) as a function of the product of the transverse magnetic field $B_{y}$ and the displacement $\Delta x$. The solid line represents a linear fit to the data points. The black dashed line indicates the reference amplitude $\Delta P_{\mathrm{ref}}$ with no transverse magnetic field or displacement. The gray shaded regions represent the $\pm10\%$ tolerance bands around the reference, respectively. }
  \label{fig:bxdy_colormap}
  \end{center}
\end{figure}

Finally, we plot in Fig.~\ref{fig:bxdy_colormap} the amplitude $\Delta P$ at zero detuning versus the product $B_{y}\Delta x$, showing, of course, a very linear dependence, with all the spectral amplitudes collapsing to a single line.  
From plots such as these, we can estimate that in order to reduce the effect of $V_m$ to less than 10\% of the $V_{\rm pnc}$ interaction, we must be able to reduce the product $|B_{y}\Delta x|$ (similarly the product $|B_{x}\Delta y|$) to a value less than $0.6 \times 10^{-11}$ Gm. In other words, for a transverse static magnetic field of $0.1$ mG, the displacement of the atomic beam from the cavity axis must be less than $60$ nm. This result provides a range within which we can adjust the magnetic field and transverse displacement independently according to the experimental requirements.  

\begin{figure}[t!]
  \begin{center}
     \includegraphics[width=0.45\textwidth]{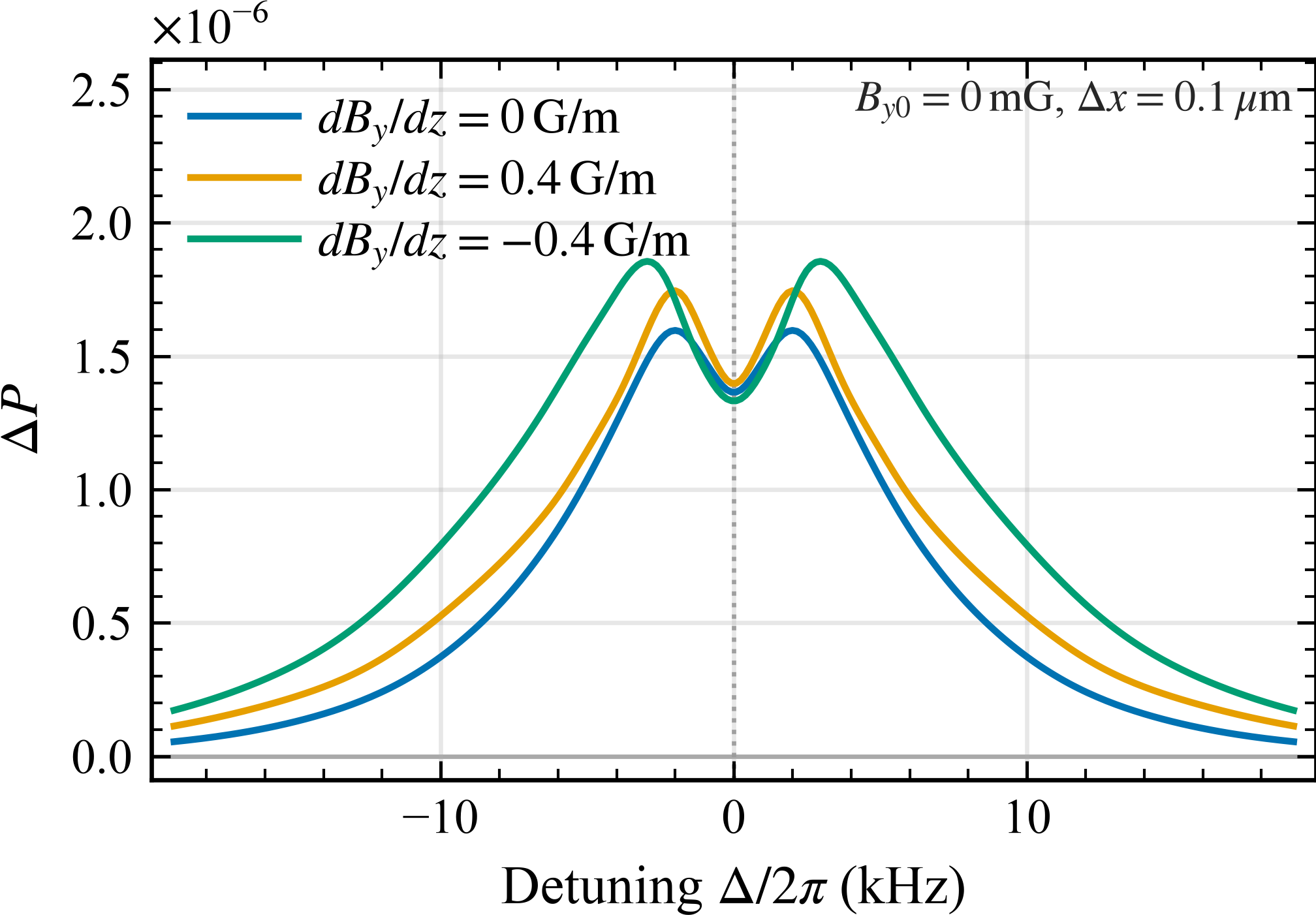}
  \caption{Spectra of the population modulation amplitude $\Delta P$ for different gradients of the $B_y$ field, where $B_{y0}$ is the magnetic field at the center of the cavity.}
  \label{fig:gradient}
  \end{center}\ 
\end{figure}

For each of the simulations discussed above, we treated the static magnetic field as if it were perfectly uniform throughout the Raman region and rf cavity.  In practice, however, due to various magnetic field coils inside the vacuum chamber, or even objects or materials outside the chamber, $B_x$ and $B_y$ may exhibit spatial gradients. We simulated the effects for different magnetic field gradients on the spectra, as shown in Fig.~\ref{fig:gradient}. 
A gradient of $\pm 0.4 \ \mathrm{G/m}$, much larger than any expected field gradient in the system, corresponds to a total field variation of $5$ mG across the cavity which is $1.25 \ \mathrm{cm}$ long. As shown in the Fig.~\ref{fig:gradient}, for the same $B_{y0} = 0$ at the center of the cavity and transverse displacement $\Delta x = 0.1  \, \mu \rm{m}$, different magnetic field gradients lead to noticeable changes in the spectral shape. However, the value of $\Delta P$ at zero detuning ($\Delta=0$) remains nearly unchanged.

\subsection{Explanation of the side-peaks in the Spectra $\Delta P$ versus $\Delta$}

\begin{figure}[b!]
  \begin{center}
     \includegraphics[width=0.45\textwidth]{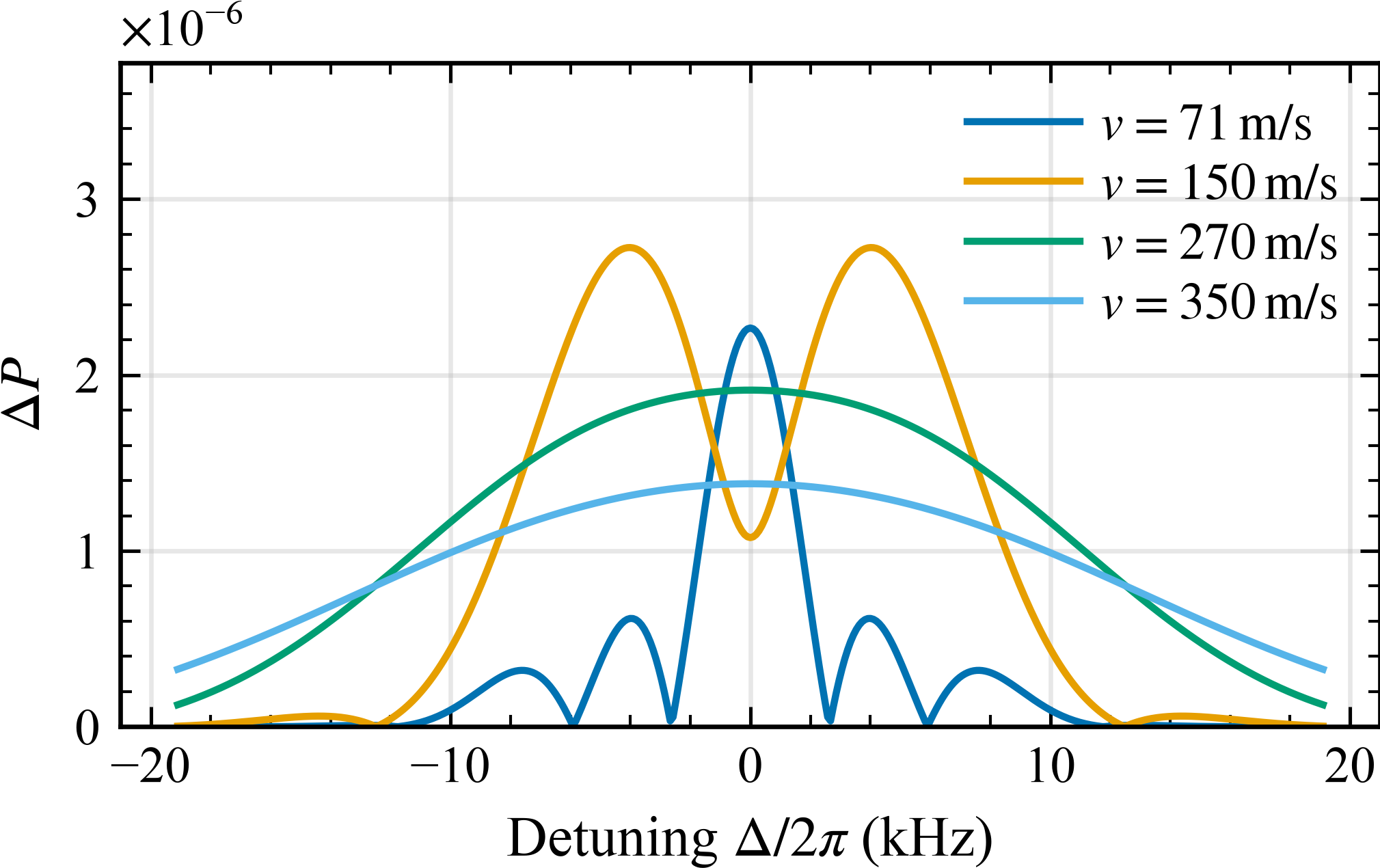}
  \caption{Plots of the amplitude $\Delta P(x=0,y=0)$ of the population modulation spectrum as a function of detuning, $\Delta/2 \pi$, for four fixed atomic velocities, $v=71,\ 150,\ 270$ and $350 \ {\rm m/s}$, for an ideal rf cavity centered at $x=0$, $y=0$, with no transverse cavity displacement.  Each spectrum is accumulated with $\sim 3000$ atoms.}
  \label{fig:fixed_speeds}
  \end{center}
\end{figure}
The shape of the spectra are interesting in that the maxima do not occur at line center. To understand the shape of the spectra, we carried out similar simulations in which we limited the width of the velocity distribution of the atoms in the atomic beam. We show a particular set of these spectra in Fig.~\ref{fig:fixed_speeds}. These plots show a broad featureless spectrum for each of the two higher velocities, $v_z = 350$ m/s and $270$ m/s. At these high velocities, the atoms spend only $\tau = d/v = 36 \ \mu$s and $46 \ \mu$s, respectively, in the rf cavity. In this limited interaction time, the Bloch precession angle is also limited to small angles.  For the lower velocities, however, the Bloch vector has more time to precess, and features start to appear in the spectra.  For example at $v_z = 150$ m/s, a pair of sidelobes appear in the spectrum near $\Delta/2 \pi = 4$ kHz. The amplitude of these sidelobes is greater than the amplitudes of the spectra for higher velocity atoms, and appear to be the source of the peaks in the spectra computed for the broad thermal velocity distribution of the full beam.  Note that in the velocity distribution function of Fig.~\ref{fig:vz_distribution}, $f(v)$ at $v_z = 150$ m/s is about half the peak value, so these atoms contribute strongly to the total spectrum.
At even lower velocity, $v_z = 71$ m/s, the interaction time is $83 \ \mu$s, and with a Rabi frequency of $\sqrt{\Omega_0^2 + \Delta^2} \approx \Delta$ when $\Delta \gg \Omega_0$, the Bloch precession angle is $\Theta_{\rm rf} = \Delta \tau \approx  \pi $ and $2 \pi$ for the two minima at 
$\Delta/2 \pi = 2.5$ kHz and $\Delta/2 \pi = 6$ kHz. When the Bloch vector precesses through a half or a whole cycle, the net population amplitude is unchanged, leading to a null in the plot. At $v_z = 71$ m/s, the distribution function $f(v)$ is quite small, so these atoms have minimal influence on the overall spectrum, but this spectrum does give us some insight into the lineshape.

\subsection{Angular Tilt of Cavity}
We have also explored the effect that the tilt of the rf cavity has on the modulation amplitude of the population, $\Delta P$. In the simulation, the rotation angle $\theta_y$ is defined as a rotation of the rf cavity about the $y-$axis, corresponding to a rotation in the \(xz\)-plane,
while the atomic trajectory remains fixed. Consequently, atoms experience a continuously varying transverse displacement in $x$ as they travel through the cavity. As the rf cavity changes orientation, the evolution of the atomic system changes due to two effects; through the change in rf-field amplitudes, and through the rotation of rf-field components from one frame into another.

\begin{figure}[b!]
  \begin{center}
     \includegraphics[width=0.45\textwidth]{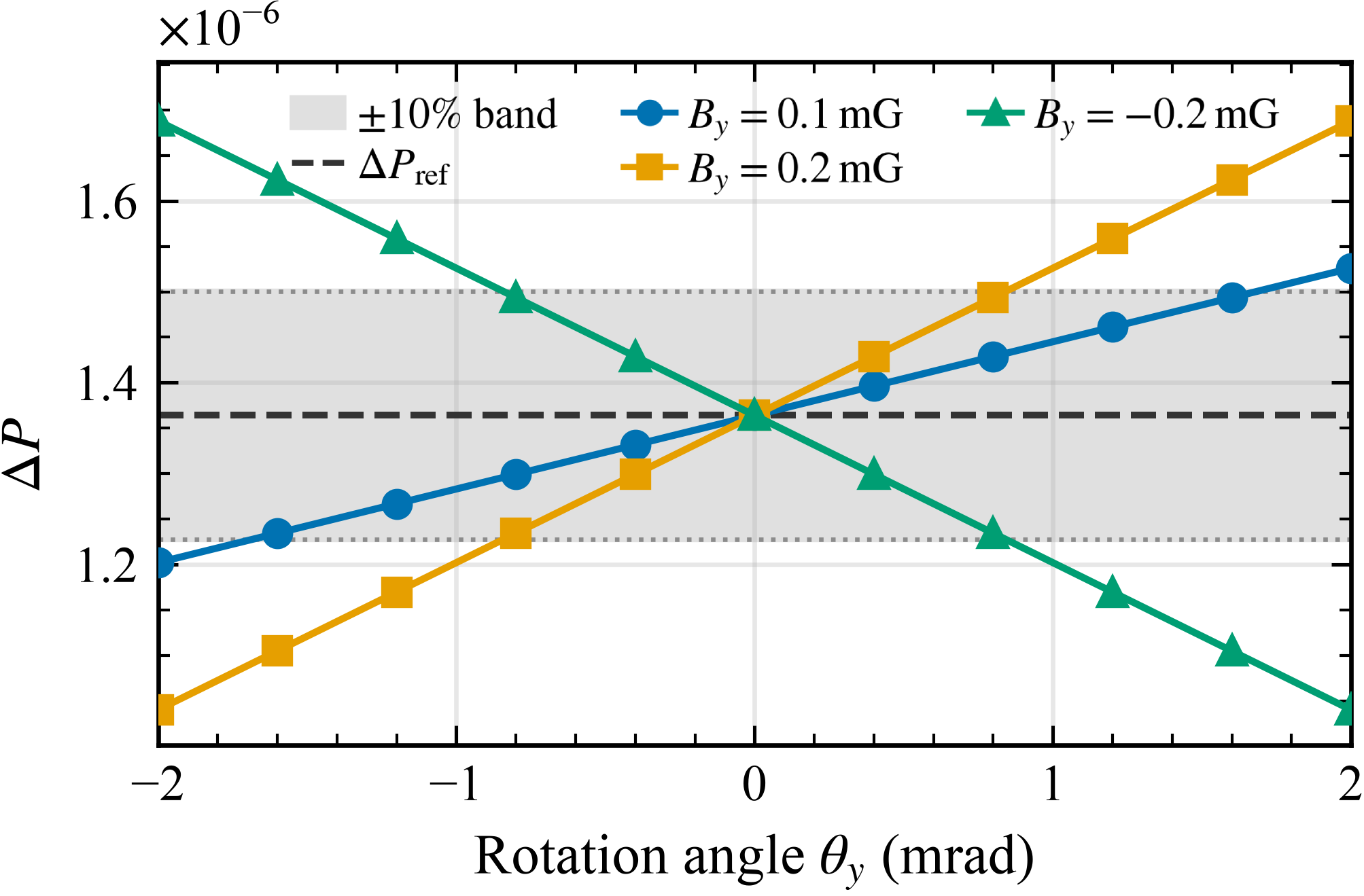}
  \caption{ Population modulation amplitude $\Delta P$ at zero detuning ($\Delta=0$) as a function of the cavity rotation angles for three different transverse magnetic fields. The solid lines are linear fits to the data points, showing the dependence of $\Delta P$ on the rotation angles for different transverse magnetic fields. The black dashed line indicates the reference amplitude, while the gray shaded region represents the $\pm 10\%$ tolerance band around the reference. }
  \label{fig:DelP_vs_tilt}
  \end{center}
\end{figure}

We show the dependence of $\Delta P$ on the cavity rotation angle $\theta_y$ at zero detuning ($\Delta=0$) for different transverse magnetic fields $B_y$ in Fig.~\ref{fig:DelP_vs_tilt}. 
All three curves intersect near $\theta_x=0$ at $\Delta P\approx1.36\times10^{-6}$. The modulation amplitude $\Delta P$ shows an approximately linear dependence on $\theta_y$ with the slope dependent on $B_y$. In particular, reversing the sign of $B_y$ from $+0.2$ to $-0.2 \ \mathrm{mG}$ reverses the slope of the angular response, while reducing the magnitude of $B_y$ weakens the angular dependence. This indicates that the sensitivity of $\Delta P$ to cavity misalignment is closely related to the magnitude of the magnetic field. Therefore, if we can reduce $B_y$ to below $0.1 \ \mathrm{mG}$, the cavity rotation angle $\theta_y$ must be kept below approximately $1.6\, \mathrm{mrad}$. Thus for an rf cavity with a length of $1.25\ \mathrm{cm}$, this requires that the transverse displacement of the cesium atomic beam between the entrance and exit of the cavity must be less than $\approx10\,\mu\mathrm{m}$. 

Since the displacement induced change in rf field experienced by the atoms occurs in the horizontal plane, the gravitational effects are symmetric with respect to positive and negative rotation angles and contribute equally to both. The effect of gravity results in a shift of the intersection point of the lines corresponding to different $B_x$ values by $0.06 \, \mathrm{mrad}$ in the positive $\theta_x$ direction. The results show that controlling the angular tilt of the cavity is as important as reducing the transverse magnetic field in the cavity in order to minimize changes in $\Delta P$, the population modulation amplitude.


\subsection{Effect of the AC Stark shift by the Raman beams  }

The spectra that we have collected to this point in the laboratory are much more complex than the relatively simple spectra calculated here.  Those spectra typically display multiple peaks of uneven spacing.  One of our goals in carrying out these numerical simulations has been to identify possible causes for the complexity of the spectra.  In this section, we explore the role of the ac Stark shift of the ground state transition by the Raman coupling beams that we use to create the 50:50 superposition state before the atoms enter the rf cavity. 
In this Raman process, the two ground states may experience different ac Stark shifts, resulting in a shift of the transition frequency in the Raman interaction region. To include the ac Stark effect in the numerical simulations, we treat the differential shift as an additional detuning of the Raman transition. The effective detuning in the Raman interaction beams is therefore given by 
\begin{equation}\label{eq:effdetuning}
    \Delta_{\mathrm{eff}}=\Delta+ \Delta_\mathrm{Stark}.
\end{equation}
where $\Delta_{\mathrm{Stark}}=2\pi\Delta\nu_{\mathrm{Stark}}= 2\pi \left(\frac{1}{2} \alpha_0 E^2\right)$, where $E$ is the Raman laser electric field amplitude and $\alpha_0$ is the scalar polarizability.  
For a spatially varying Raman field, the Stark shift follows the local Raman laser profile and is included in the phase accumulated during the Raman interaction. As the atom moves through the Raman beam, the Stark-induced detuning changes with position. The corresponding phase accumulated during the Raman interaction is
\begin{equation}\label{eq:accumphase}
\phi(t) = \int_0^t\Delta_{\mathrm{eff}}(t')dt',
\end{equation}
and the phase factor $e^{i\phi(t)}$ is used in the Raman evolution. The evolution of the state vector in the rf cavity is not otherwise modified by the ac Stark shift in the simulation.

We calculated the ac Stark polarizability of the cesium ground-state hyperfine levels using the frequency-dependent polarizability expression given by Milroy, Safronova, and Clark~\cite{mitroy2010theory}, which accounts for the contributions of intermediate electric-dipole-coupled states. For a ground state $\lvert\psi\rangle$, the frequency-dependent scalar polarizability is given by
\begin{equation}\label{eq:polarizability}
\alpha_0 =
\frac{2}{3(2J'+1)}
\sum_n
\frac{
\Delta E_n
\left|
\left\langle
\psi \left\| D \right\| \psi_n
\right\rangle
\right|^2
}{
(\Delta E_n)^2-\omega^2
},
\end{equation}
where $J'$ is the electronic angular momentum of the ground state, $\Delta E_n=E_n-E_g$ is the energy separation between the ground state and the intermediate state $\lvert\psi_n\rangle$, and $D=rC^{(1)}(\hat{\mathbf r})$ is the electric-dipole operator, with $C^{(1)}$ representing angular factor.  
The Stark shift of the $F=3$ ground state results from dipole coupling to the $F=2$-4 hyperfine components of the $6p_{3/2}$ level.  For the Stark shift of the $F=4$ ground state, we include terms from the $F=3$-5 components.  (This is in contrast to the Raman coupling, which includes only the $F=3$ and 4 intermediate states due to electric dipole selection rules.

\begin{figure}[b!]
  \begin{center}
     \includegraphics[width=0.45\textwidth]{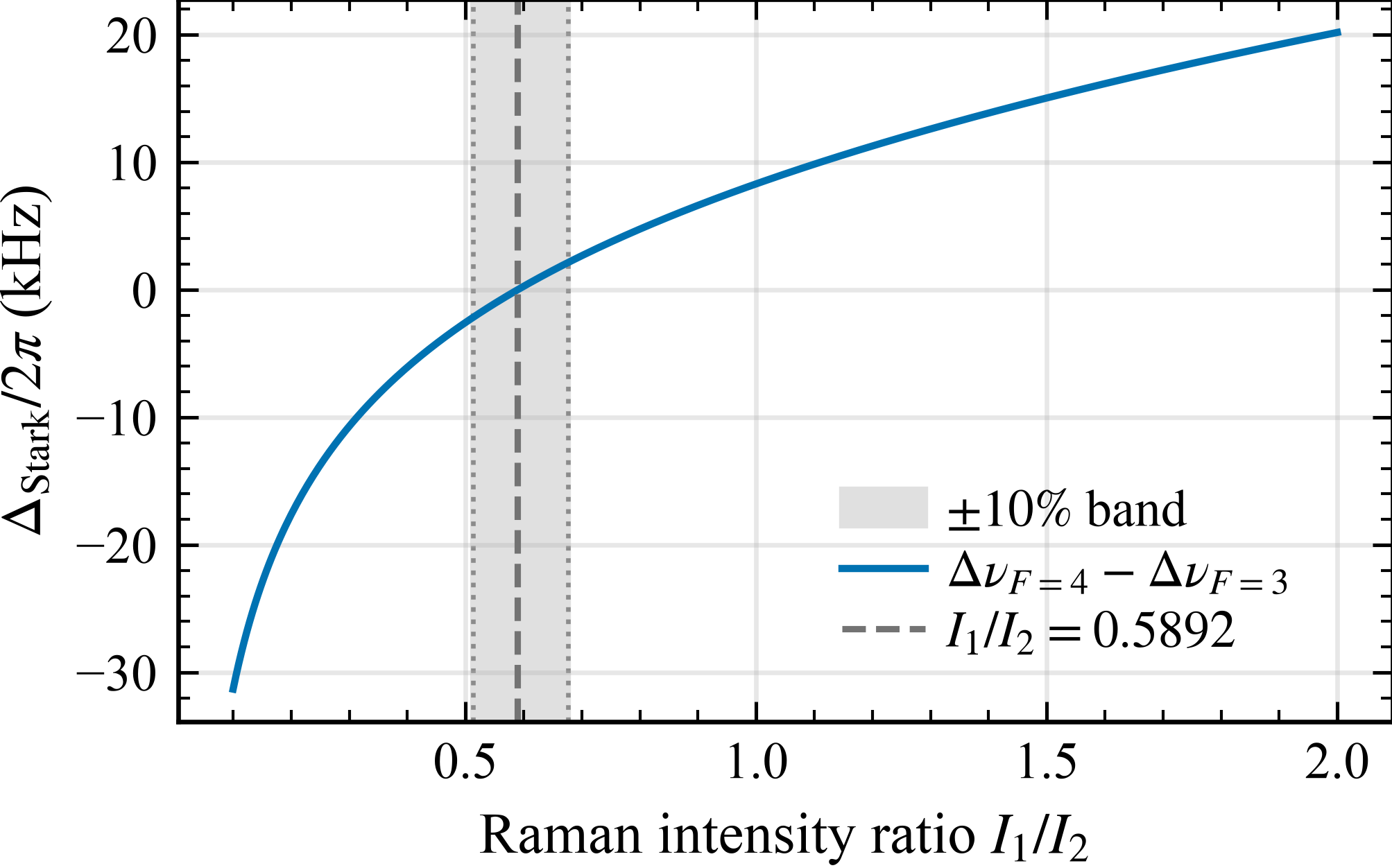}
  \caption{Differential ac Stark shift as a function of the Raman intensity ratio $I_1/I_2$. The shift vanishes near $I_1/I_2=0.5892$, where the Stark shifts of the two ground states become equal.}
  \label{fig:diffstark}
  \end{center}
\end{figure}

In Fig.~\ref{fig:diffstark}, we show the differential ac Stark shift, $\Delta\nu_{\mathrm{Stark}}=\Delta\nu_{F=4}-\Delta\nu_{F=3}$, as a function of the intensity ratio of the two Raman lasers. 
In the calculation, the geometric mean of the two laser powers is kept constant $\sqrt{P_1 P_2}=10\ \mathrm{mW}$ such that the Raman coupling strength remains approximately unchanged, while the intensity ratio $I_1/I_2=P_1/P_2$ is varied. All calculations use polarizabilities evaluated at a laser detuning of 250 MHz below the $6P_{3/2}, F'=2$ excited state. When the two Raman beams have equal intensities, the differential ac Stark shift is approximately $8 \ \mathrm{kHz}$. When $I_1/I_2$ is approximately $0.59$, while keeping $\sqrt{I_1I_2}$ constant, the differential ac Stark shift approaches zero
, while the Raman coupling strength remains unchanged. This methodology will allow us to minimize the effect of the ac Stark shift on the measured $\Delta P$.

\begin{figure}[t!]
  \begin{center}
     \includegraphics[width=0.45\textwidth]{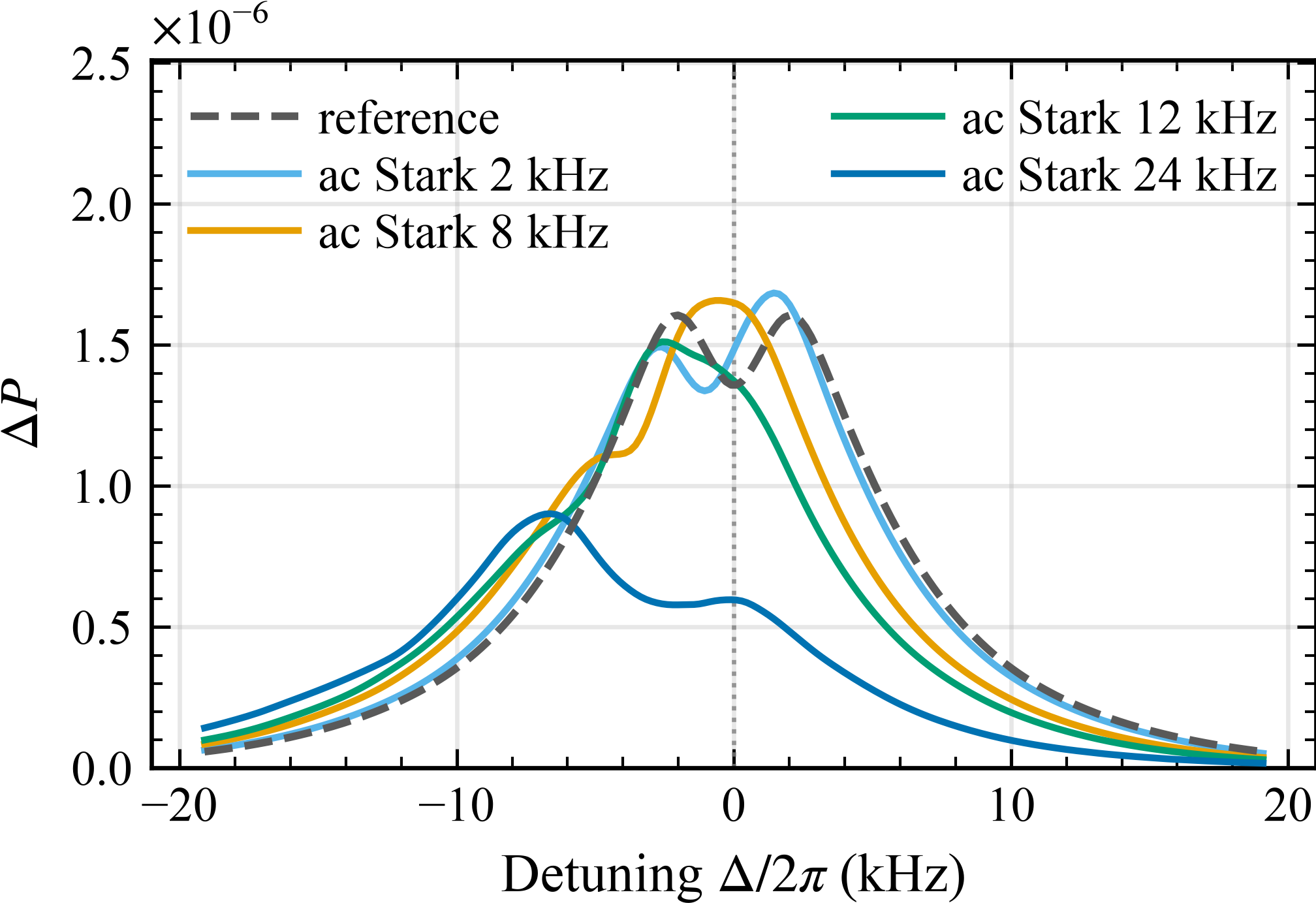}
  \caption{Raman transition spectra \(\Delta P\) calculated for different differential AC Stark shifts, ranging from $2$ to $24~\mathrm{kHz}$. The dashed curve shows the reference spectrum without the AC Stark shift.}
  \label{fig:shiftspec}
  \end{center}
\end{figure}

Fig.~\ref{fig:shiftspec} illustrates the effect of the differential ac Stark shift on the Raman transition spectrum $\Delta P$. A differential shift of $2~\mathrm{kHz}$ introduces a clear asymmetry into the double-peaked spectrum, suppressing the left peak while enhancing the right peak relative to the reference spectrum. At $8~\mathrm{kHz}$, which approximately corresponds to equal intensities of the two Raman beams, the right-hand peak becomes strongly dominant. As the differential shift is increased to $12~\mathrm{kHz}$, the double-peak structure is largely washed out and the spectral lineshape is substantially modified. At $24~\mathrm{kHz}$, the asymmetry reverses, with the left-hand peak becoming dominant, while the overall $\Delta P$ is reduced to approximately half of the reference value. Reversing the sign of the differential AC Stark shift produces the corresponding spectral distortion in the opposite direction, demonstrating the expected symmetry of the spectrum with respect to the sign of the shift.

\subsection{Future Improvements to the Experimental Setup }
As we have shown in the previous sections, we must reduce the products $B_x \Delta y$ and $B_y \Delta x$ to values less than $0.6 \times 10^{-11}$ Gm. To be able to achieve this standard, we require a method for measuring the transverse magnetic field components, and for controlling and sensing minute displacements of the rf cavity.  For the former, we have designed and fabricated a custom printed circuit board (PCB) incorporating MMC5983MA magnetometers to measure the magnetic field in the rf interaction region.
Four magnetometer chips are evenly spaced around a circular opening in each PCB. We have one PCB positioned where the atomic beam enters the rf cavity and the other where it exits the rf cavity. Both PCBs are connected using  vacuum feedthroughs to an external microcontroller unit (MCU) and display. Spatial averaging of the four sensor readings from each PCB combined with temporal averaging enable the magnetic field in the rf region to be measured with a precision of approximately $0.1$ mG. This arrangement allows the magnetic field to be monitored at both ends of the rf cavity without blocking the atomic beam. Averaging the measurements from the two PCBs provides an estimate of the magnetic field within the rf cavity, while comparing their readings provides an indication of field variations along the beam axis. We also tested the temporal stability of the magnetometers, observing a drift of $\sim 0.014$ mG/hour or smaller over a period of 16 hours.  Repeated measurements also enabled slow temporal drifts to be tracked, providing a practical reference for adjusting the magnetic compensation field and assessing the stability of the magnetic environment during operation. 

For accurate positioning of the cavity, we have mounted the cavity on a five-axis positioner driven by a pico-motor translator, capable of step sizes of 30 nm.  Our current positioner is not sufficiently repeatable, so we have been exploring a position sensor based on potentiometers coupled to the motors. We also note that a number of piezoelectric stages with sub\JQ{-}nanometer resolution and repeatability are commercially available. In our next round of laboratory efforts, we will be implementing one of these various techniques. 

\section{Conclusions}
Through the analyses presented in this paper, we can make several conclusions that will help guide future efforts in the laboratory.  First, we have shown that controlling the products $B_x \Delta y$ and $B_y \Delta x$ to values less than $0.6 \times 10^{-11}$ Gm will allow us to reduce the magnetic dipole interactions to less than 10\% of the E1 moment that we seek to measure.  Furthermore, the plots of Fig.~\ref{fig:amp_vs_By} suggest that one can reduce $B_y$ by varying $\Delta x$ until little variation in the spectrum is observed.  The same procedure may be followed to reduced $B_x$, searching for $\Delta y$ at which the spectrum is insensitive to further change.  When $B_x$ and $B_y$ are minimized, then the spectrum should also be insensitive to angular tilts of the rf cavity.  Perhaps the most significant result of the studies presented here are related to the ac Stark shift of the Raman beams, and how this shift varies with the relative intensity of the two Raman beams.  One potential approach, which can be implemented through selecting the correct relative intensity for the two Raman beams by imposing phase modulation sidebands on a single laser beam such that the carrier serves as one Raman beam and an optical sideband as the other, is an attractive solution and relatively easily implemented. 

\section{Data Availability Statement}
The data that support the findings of this article are openly available~\cite{PURR5294}. 

\section{AI Use Disclosure}
The xAI Grok 4.6 and OpenAI GPT-6 models were used to assist with modifications to Python scripts for parameter scans and visualization of numerical results. The physical models, core numerical algorithms, and main simulation code were developed by the authors. All AI-assisted code modifications and resulting figures were reviewed and verified by the authors, who take full responsibility for the reported results.

\section{Acknowledgment}\label{sec:Acknowledgment}
This material is based upon work supported by the National Science Foundation under Grant Number 2409461-PHY.  We also acknowledge technical assistance with magnetic field sensor circuits by S. Hokanson.


\bibliography{biblio}

\end{document}